# Temperature-programmed reduction and dispersive X-ray absorption spectroscopy studies of $CeO_2$-based nanopowders for intermediate-temperature Solid-Oxide Fuel Cell anodes

*Marina S. Bellora*[1,2], *Joaquín Sacanell*[2,3,]*, *Cristián Huck-Iriart*[1], *Analía L. Soldati*[3], *Susana A. Larrondo*[4,5], *and Diego G. Lamas*[1]

[1] CONICET / Laboratorio de Cristalografía Aplicada, Escuela de Ciencia y Tecnología, Universidad Nacional de General San Martín, San Martín, Argentina

[2] Departamento de Física de la Materia Condensada, Gerencia de Investigación y Aplicaciones, Centro Atómico Constituyentes, Comisión Nacional de Energía Atómica, San Martín, Argentina

[3] Instituto de Nanociencia y Nanotecnología (INN), CNEA-CONICET, San Martín and San Carlos de Bariloche, Argentina

[4] UNIDEF-CONICET-MINDEF, Departamento de Investigaciones en Sólidos, CITEDEF, Villa Martelli, Argentina

[5] Instituto de Investigación e Ingeniería Ambiental, Universidad Nacional de General San Martín, San Martín, Argentina

**Abstract**

In this work, we study the influence of the average crystallite size and dopant oxide on the reducibility of $CeO_2$-based nanomaterials. Samples were prepared from commercial $Gd_2O_3$-, $Sm_2O_3$- and $Y_2O_3$-doped $CeO_2$ powders by calcination at different temperatures ranging between 400 and 900ºC and characterized by X-ray powder diffraction, transmission electron microscopy and BET specific surface area. The reducibility of the samples was analyzed by temperature-programmed reduction and *in situ* dispersive X-ray absorption spectroscopy techniques. Our results clearly demonstrate that samples treated at lower temperatures, of smallest average crystallite size and highest specific surface areas, exhibit the best performance, while $Gd_2O_3$-doped ceria materials display higher reducibility than $Sm_2O_3$- and $Y_2O_3$-doped $CeO_2$.

*Corresponding author:

Joaquín Sacanell

E-mail: sacanell@tandar.cnea.gov.ar

## Introduction

Cerium oxide ($CeO_2$) has been studied for many years and it is still the focus of great attention due its wide range of possible applications. It is used in three-way catalysts (TWCs) for the elimination of toxic auto-exhaust gases [1,2], low-temperature water-gas shift (WGS) reaction [3,4], oxygen sensors [5,6], oxygen permeation membrane systems [7,8] and solid-oxide fuel cells (SOFCs) [9,10,11,12], among others.

In all the aforementioned applications, there is a growing interest in the study of nanostructured materials, in which an improved surface-to-volume ratio can be obtained. This interest is founded in the consequent substantial reduction in the energy for defect formation occurs in nanocrystalline $CeO_2$, that leads to a high degree of non-stoichiometry and electronic carrier generation [13] both beneficial for redox, catalytic and transport properties. For example, $CeO_2$-based nanoceramics exhibit enhanced ionic conductivity [14], which is very important for their application as SOFC electrolytes.

Although the benefit of the use of nanostructured materials has already been pointed out by several authors, the studies of this type of materials from the basic point of view to understand the mechanisms underlying the different processes of catalysis are scarce. In recent years, our research groups have investigated $CeO_2$-based and $NiO/CeO_2$-based catalysts by dispersive X-ray absorption spectroscopy (DXAS) under different atmospheres and reaction conditions [15,16,17,18]. Samples with different morphologies were analyzed, finding that $CeO_2$-based materials with small average crystallite size and high specific surface area exhibit the best properties, reaching high reducibility and excellent methane conversion for intermediate temperatures [19].

In this work, we analyzed the redox behavior of nanocrystalline $Gd_2O_3$, $Sm_2O_3$ and $Y_2O_3$-doped $CeO_2$ (GDC, SDC and YDC, respectively) powders. We used conventional laboratory temperature programmed reduction (TPR) and combined those results with an *in-situ* DXAS study under diluted hydrogen atmosphere, in order to gain further insight on the redox properties of the systems. The samples were treated at different temperatures, between 400 and 900°C to analyze the influence of the crystallite size.

## Experimental procedure

Samples were prepared from nanocrystalline commercial (Nextech Materials) $Ce_{0.8}Gd_{0.2}O_{1.9}$ (GDC), $CeO_2$-10 mol% $Sm_2O_3$ ($Ce_{0.82}Sm_{0.18}O_{1.91}$, SDC) and $CeO_2$-10 mol% $Y_2O_3$ ($Ce_{0.82}Y_{0.18}O_{1.91}$, YDC) powders. They were calcined at 400°C, 650ºC and 900ºC in order to study the influence of the average crystallite size.

X-ray powder diffraction (XPD) was performed in a Brucker D8 Discover DaVinci diffractometer (Institute of Physics, University of Sao Paulo, Brazil) operated with Cu-K$\alpha$ radiation at 40 kV and 30 mA, a Ni filter and a Lynx-eye detector, in Bragg-Brentano configuration. Experimental data were collected in the angular 2$\theta$ range of 20-140° with a step size of 0.02° and a time per step of 1 s. The average crystallite sizes of the crystalline phases were determined for all the samples using the Scherrer equation.

Specific surface area was evaluated by means of $N_2$-physisorption with a Quantachrome Corporation Autosorb-1 equipment. Samples were previously degassed with pure He at 90°C during 12 h. Results were obtained using the five-point Brunauer−Emmett−Teller (BET) method.

Transmission Electron Microscopy (TEM) experiments were performed using a Philips CM 200 UT microscope operated at 200kV. The microscope was equipped with ultratwin objective lens and an EDAX spectrometer for chemical analysis by EDS. Powdered samples were suspended in isopropyl alcohol, ultrasonicated for 2 minutes and deposited in Cu/ultrathin hollow carbon TEM grids (Ted Pella).

Hydrogen temperature programmed reduction (TPR) experiments were performed in a Micromeritics Chemisorb 2720 equipment to study sample reducibility. The mass employed for each experiment was of 80 mg. Prior to TPR tests, samples were pretreated in He at 300°C during 30 min to remove any adsorbed species on the solid surface. TPR was carried out with a 50 $cm^3$(STP) $min^{-1}$ (5 vol.% $H_2$/Ar) flow from room temperature up to 800 °C following a heating ramp of 10 °C $min^{-1}$. Hydrogen uptake was estimated using a Thermal Conductivity Detector (TCD) previously calibrated.

Dispersive X-ray absorption spectroscopy (DAXS) study was performed at the D06A-DXAS dispersive beamline of the Brazilian Synchrotron Light Laboratory (LNLS, Campinas, Brazil). A Si (111) monochromator was used altogether with a CCD detector to collect the absorption spectrum in transmission mode. Self-supporting discs were prepared by mixing the sample powder with boron nitride that has no significant absorption in the energy ranges used. The catalyst mass in the discs was calculated in order to obtain a total absorption ratio of 1.5. Sample discs were located in a sample-holder with a thermocouple attached to it. The sample holder was placed in a quartz reactor, with inlet and outlet gas lines, and located in a furnace with temperature control. Inlet gas composition was set with a gas-mixing station provided with mass flow controllers and exit composition was assessed with a Pfeiffer Omnistar mass spectrometer.

We followed the evolution of DXAS spectra as function of temperature at the $L_3$-edge of Ce under diluted $H_2$(5% in He), in order to determine the $Ce^{3+}/Ce^{4+}$ proportion. The temperature range was of 400-800°C. $CeO_2$ and $Ce(NO_3)_3.6H_2O$ were also measured as standards. $Ce^{3+}/Ce^{4+}$ proportion was obtained by means of linear least square procedures using Python scripting.

## Results and discussion

X-ray diffraction patterns for the samples calcined at 400ºC are presented in Figure 1. All of them display the expected fluorite-type crystal structure with no sign of any impurity. The nanostructured character of the powders is evidenced by their wide Bragg peaks.

Samples of the three compounds were treated at different temperatures, in order to obtain powders with different particle sizes. Figure 2 displays the X-ray diffraction patterns for YDC nanopowders treated at 400ºC, 650ºC and 900ºC. As expected, the broadening of Bragg peaks decreases for increasing calcination temperature because samples treated at higher temperatures are formed by larger crystallites. The average crystallite sizes (D) determined from the Scherrer equation for all samples are summarized in Table I. Crystallite sizes range from around 5 to ~50 nm. In particular, the SDC sample treated at 900ºC, is formed by crystallite significantly larger than GDC and YDC samples calcined at the same temperature.

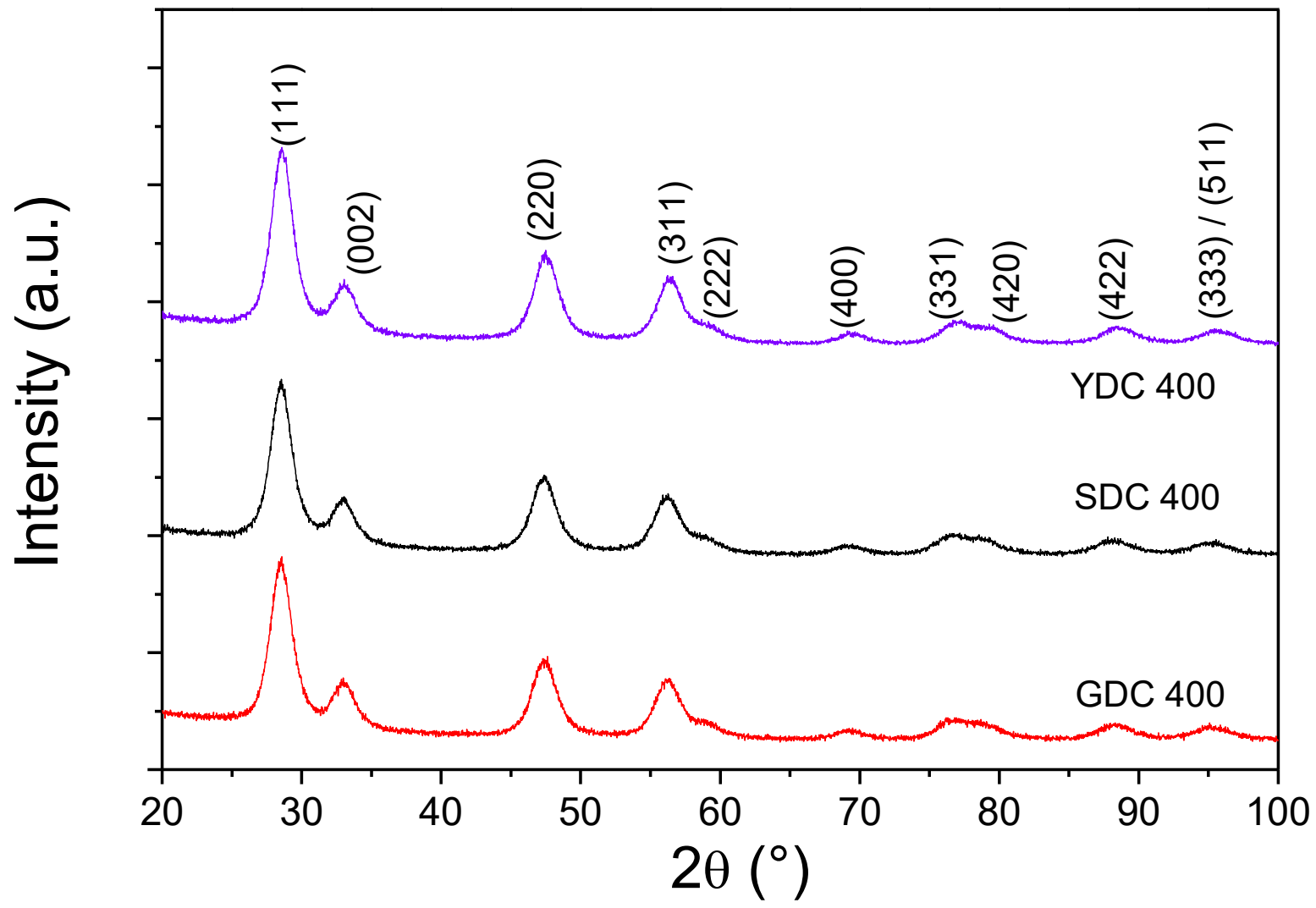

**Figure 1:** X-ray diffraction patterns for all compounds treated at 400ºC.

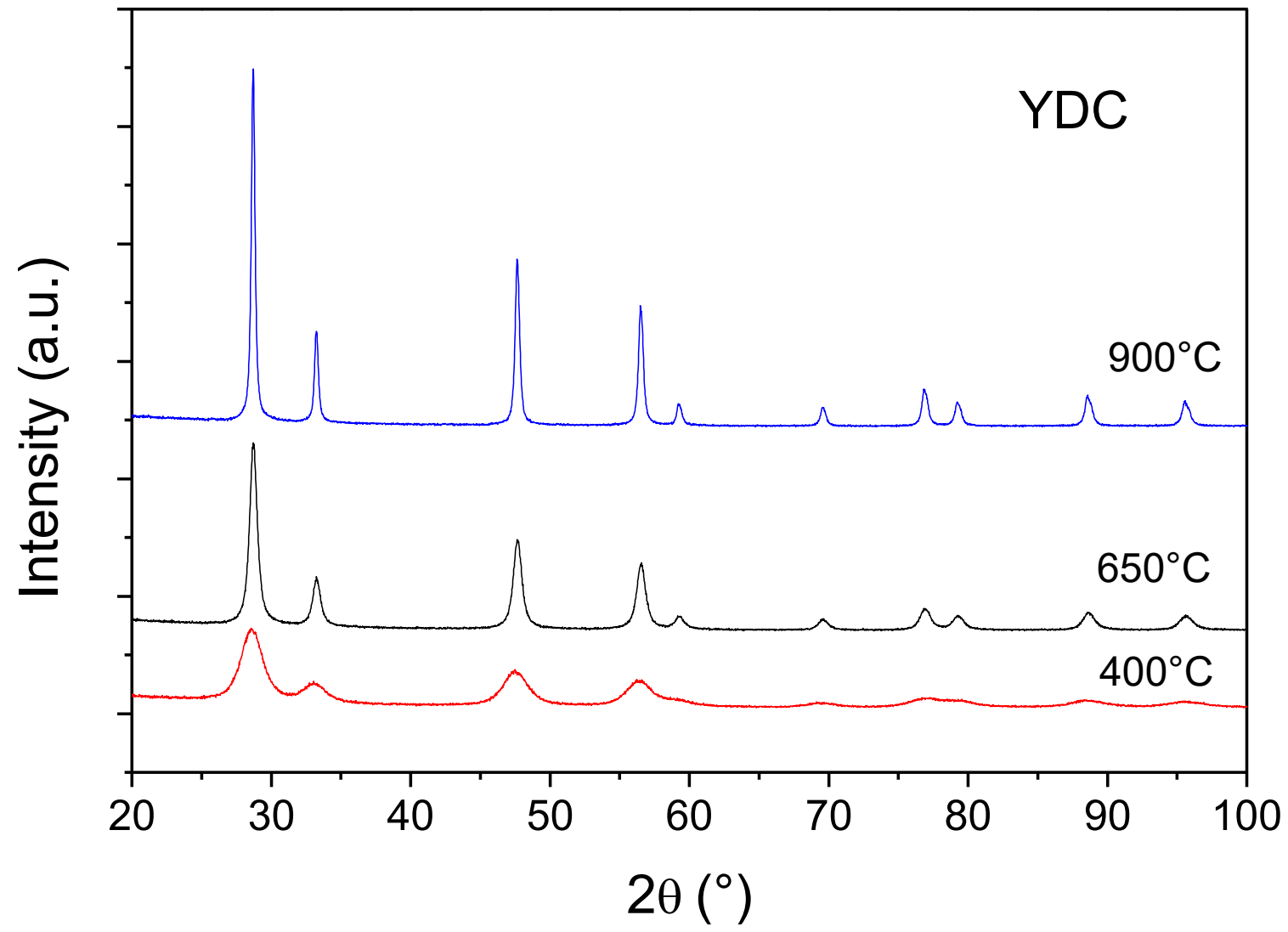

**Figure 2:** X-ray diffraction patterns for YDC powders compounds treated at 400ºC, 650ºC and 900ºC.

| Sample | 400ºC | 650ºC | 900ºC |
|---|---|---|---|
| **GDC** | 4.5(2) | 14(1) | 36(3) |
| **SDC** | 4.6(2) | 13(1) | 48(4) |
| **YDC** | 4.6(2) | 14(1) | 36(3) |

**Table I:** Average crystallite size in nm determined using Scherrer equation for all the samples studied in this work.

Figure 3 displays selected TEM images taken from several samples treated at 650ºC. From those images, it can be observed that particle sizes are of the same order of magnitude of crystallite sizes (Table I), thus indicating that particles are single crystals. This feature was also observed for samples treated at 400ºC and for samples treated at 900ºC, as exemplified in Figure 4 for the SDC system. From the morphological point of view, it can be seen that the particles display polygonal shapes in samples treated at higher temperatures.

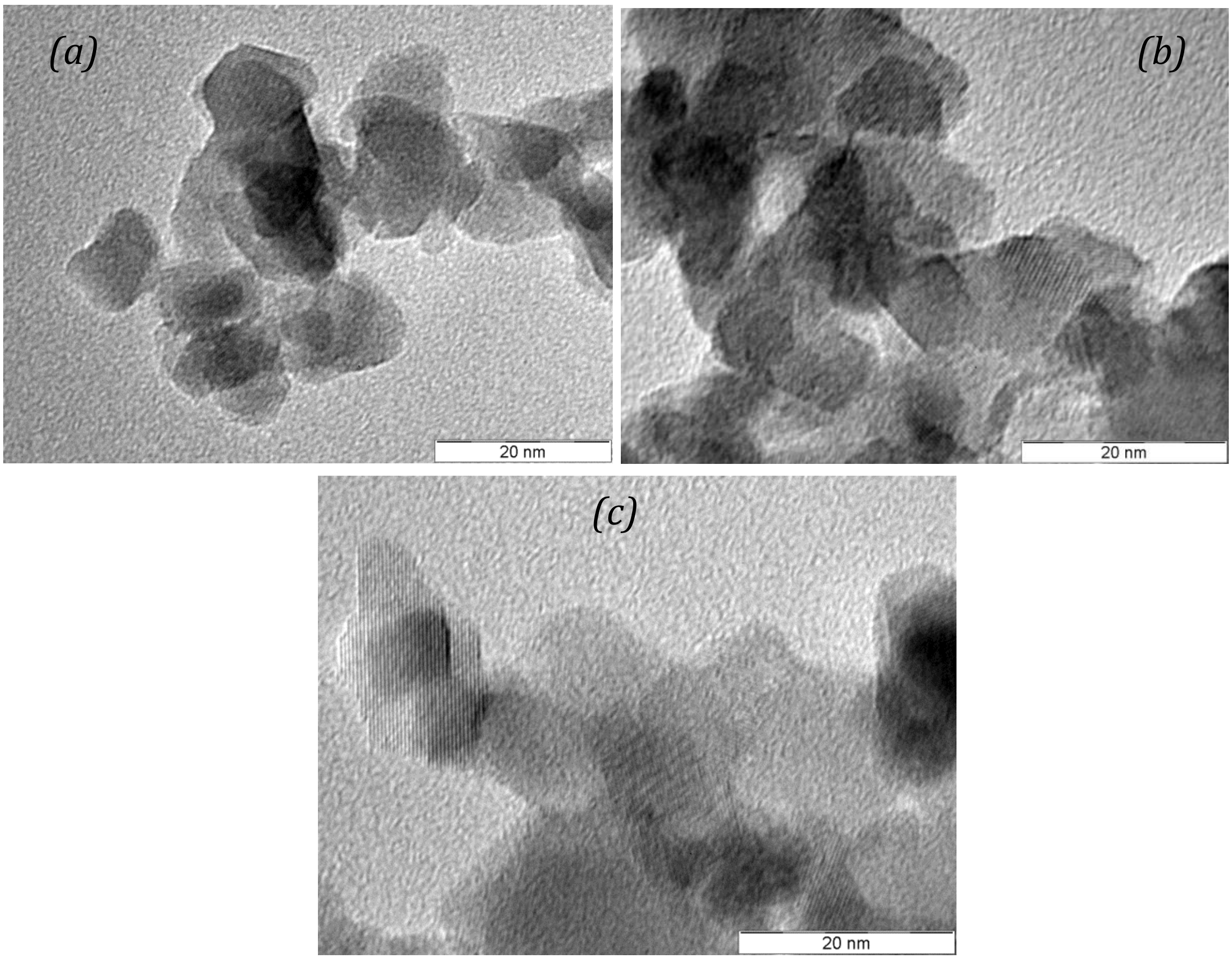

**Figure 3:** TEM images of the samples calcined at 650°C: (a) GDC, (b) SDC, (c) YDC.

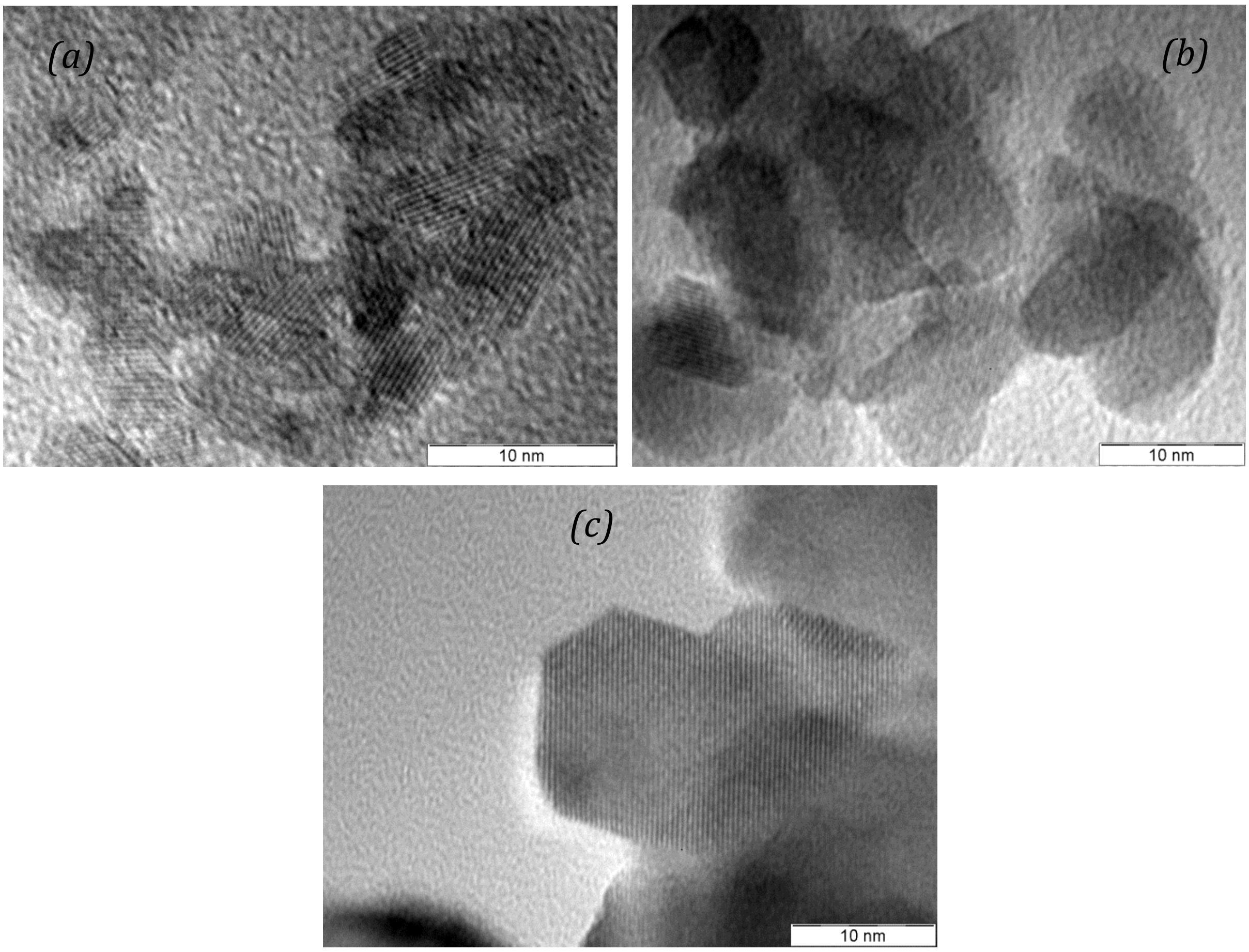

**Figure 4:** TEM images for the SDC samples treated at (a) 400°C, b) 650°C, c) 900ºC.

The dependence of the average crystallite size as a function of calcination temperature is presented in Figure 5, where it can be seen that crystallite growth is less significant in GDC and YDC compared to SDC. In any case, in undoped ceria the effect is more prominent than that observed in the present work [20,21]. This is a consequence of the inhibition produced by the solute drag model due to a space charge effect [22]. $Y_2O_3$ is known as one of the most efficient inhibitors of grain boundary mobility [20], here we show that $Gd_2O_3$ has a very similar effect.

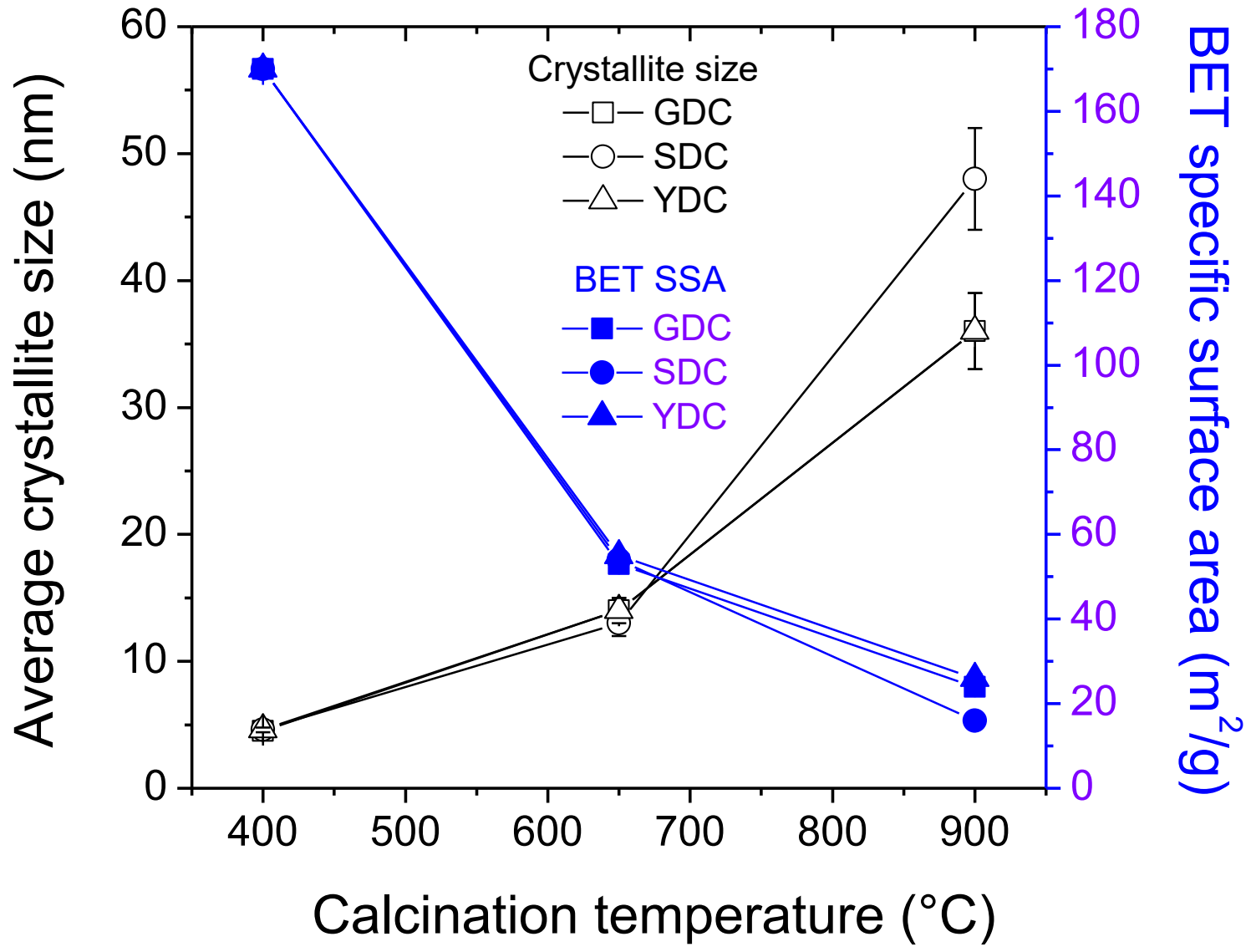

**Figure 5:** Average crystallite size and BET specific surface area as functions of calcination temperature.

The values of BET specific surface area of the samples are summarized in Table II. As it can be observed in Figure 5, there is a marked correlation between SSA and average crystallite size, showing inverse trends for increasing calcination temperature.

| Sample | 400°C | 650°C | 900°C |
|---|---|---|---|
| **GDC** | 170 | 53 | 24 |
| **SDC** | 170 | 54 | 16 |
| **YDC** | 170 | 55 | 26 |

**Table II:** BET specific surface area in m$^2$/g for all samples.

The temperature-programmed reduction (TPR) profiles of samples calcined at 650°C are shown in Figure 6. The TPR signal, directly related with hydrogen consumption, is presented as a function of the temperature of the sample. The curves do not present the same shape, position and height of the maximum. It is clearly observed that the main peak for all samples is close to 500°C, while a secondary peak is detected around 770°C. The main peak corresponding to the GDC sample, notably occurs at a lower temperature than those corresponding to YDC and SDC, indicating that the first is more efficient for hydrogen oxidation than the other two. Besides, SDC sample is slightly better than YDC one considering their reducibility, in view of the lower temperature at which hydrogen consumption is triggered. No significant difference is observed between the secondary peak of the three samples. The same trend was observed for samples treated at 400 and 900ºC. In Figure 6(b) we show the hydrogen consumption as a function of temperature in which the overall picture is clarified. Indeed, hydrogen consumption of GDC triggers at significantly lower temperature and is higher in the whole temperature range, than that corresponding to SDC and YDC, which are both very similar.

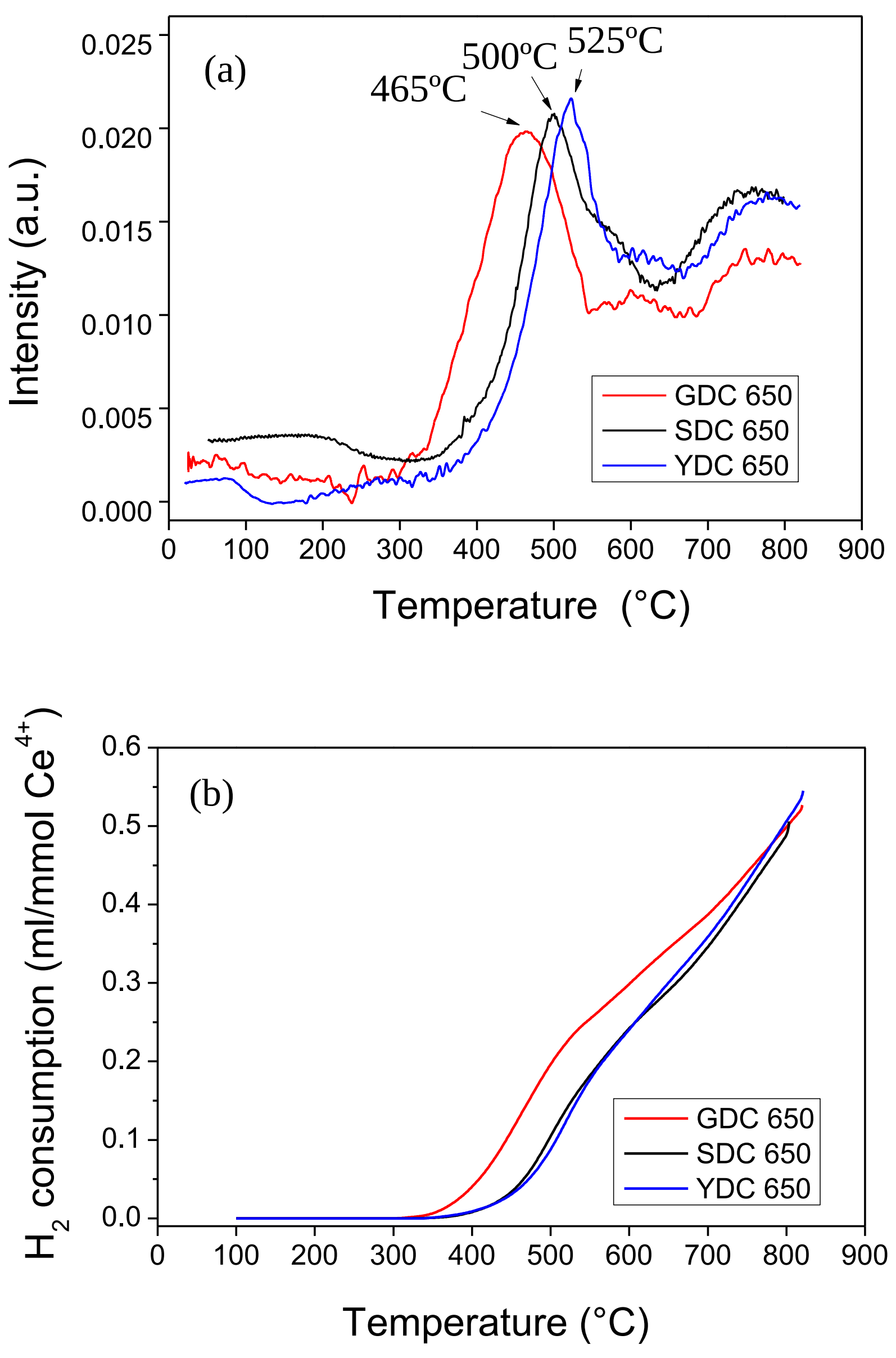

**Figure 6:** (a) TPR profile and (b) hydrogen consumption, for samples calcined at 650°C.

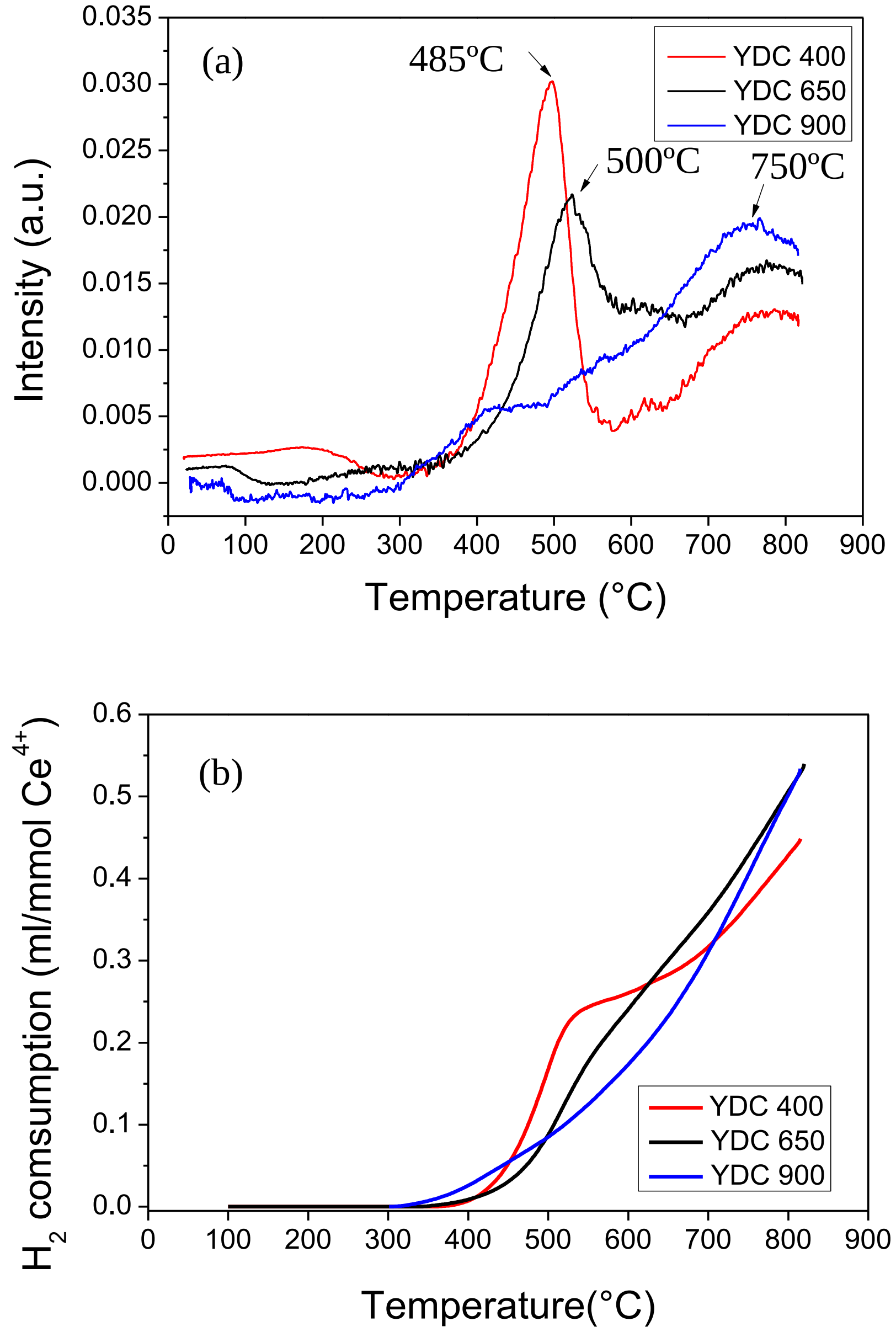

**Figure 7:** (a) TPR profiles and (b) hydrogen consumption for YDC samples.

We also studied the influence of the average crystallite size on the reducibility of the samples. Figure 7(a) displays the TPR profiles for the YDC system as an example. It can be observed that the temperature of the main peak decreases with increasing calcination temperature, thus indicating that samples of lower average crystallite size are more efficient than those with larger particles. The same overall trend is observed for the other two systems (not shown). Regarding hydrogen consumption (Figure 7(b)), it can be noticed that it is enhanced in the sample treated at 400ºC, confirming again the enhanced reducibility of samples with smallest crystallites.

The above results regarding the influence of crystallite size and dopant oxide on the reducibility of the samples were further confirmed by using the DXAS technique under diluted $H_2$ atmosphere (5 mol% $H_2$/He). The experimental procedure was designed in order to mimic the conditions of the laboratory TPR experiments. For example, DXAS data taken at the Ce $L_3$-edge for the SDC nanopowder calcined at 400°C is shown in Figure 8, displaying the evolution as a function of temperature. At room temperature, two peaks characteristic of $Ce^{4+}$ are detected, while one peak corresponding to $Ce^{3+}$ becomes more prominent at high temperatures. By measuring DXAS data corresponding to $Ce^{3+}$ and $Ce^{4+}$ standards and using a linear combination procedure, it was possible to determine the $Ce^{3+}$ fraction as a function of temperature for all the samples.

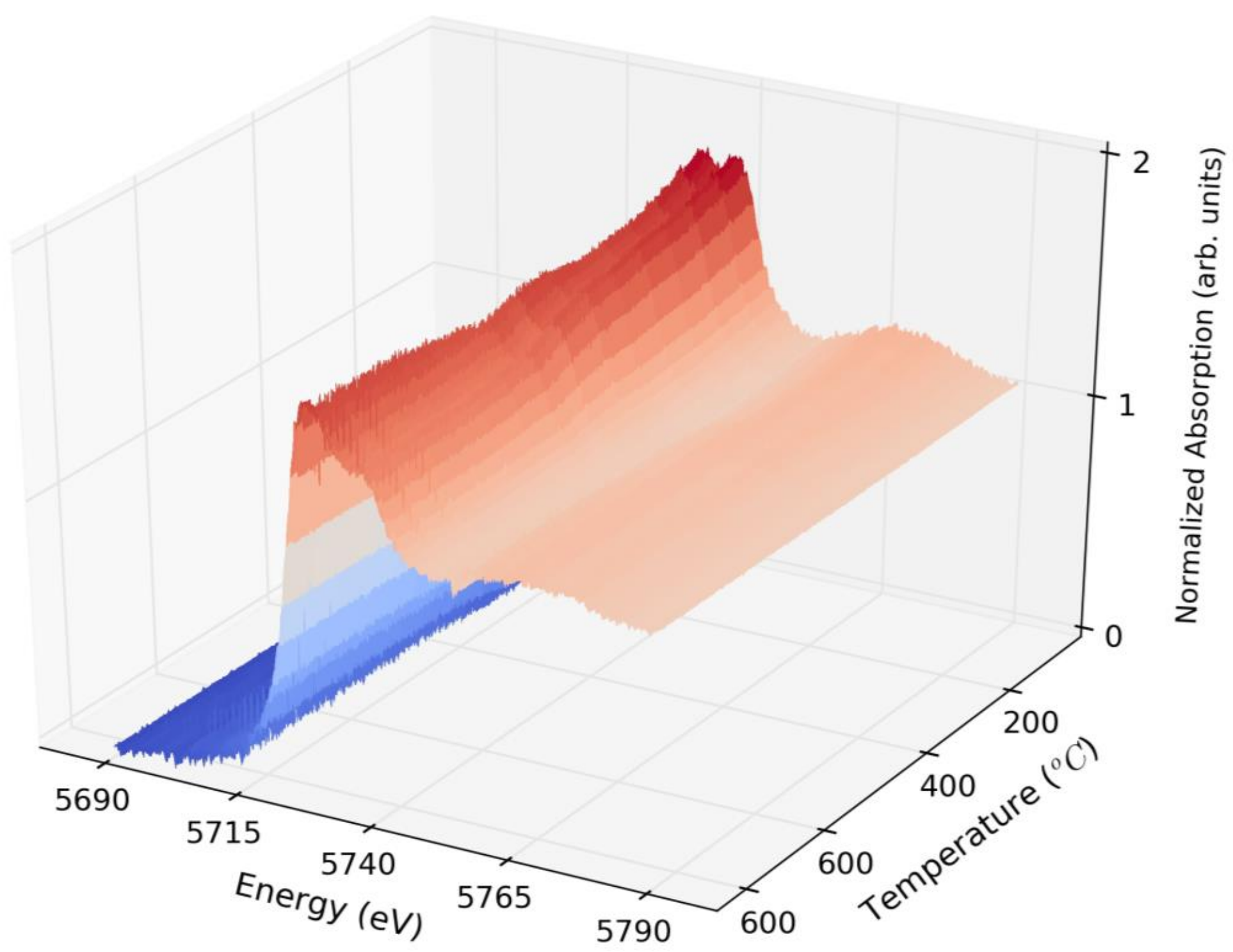

**Figure 8:** DXAS data at the Ce $L_3$-edge collected upon heating under 5 mol% $H_2$/He up to 600°C for SDC nanopowder calcined at 400°C.

Figure 9 displays the $Ce^{3+}$ fraction ($\alpha_{Ce}{}^{3+}$) as a function of temperature for GDC, SDC and YDC samples treated at 400ºC (Fig. 8(a)) and 650ºC (Fig. 8(b)). As it can be observed that the largest values of $\alpha_{Ce}{}^{3+}$ were exhibited by GDC for both sets samples (calcined at 400 and 650°C), clearly indicates its higher reducibility compared to SDC and YDC systems, in excellent agreement with TPR results.

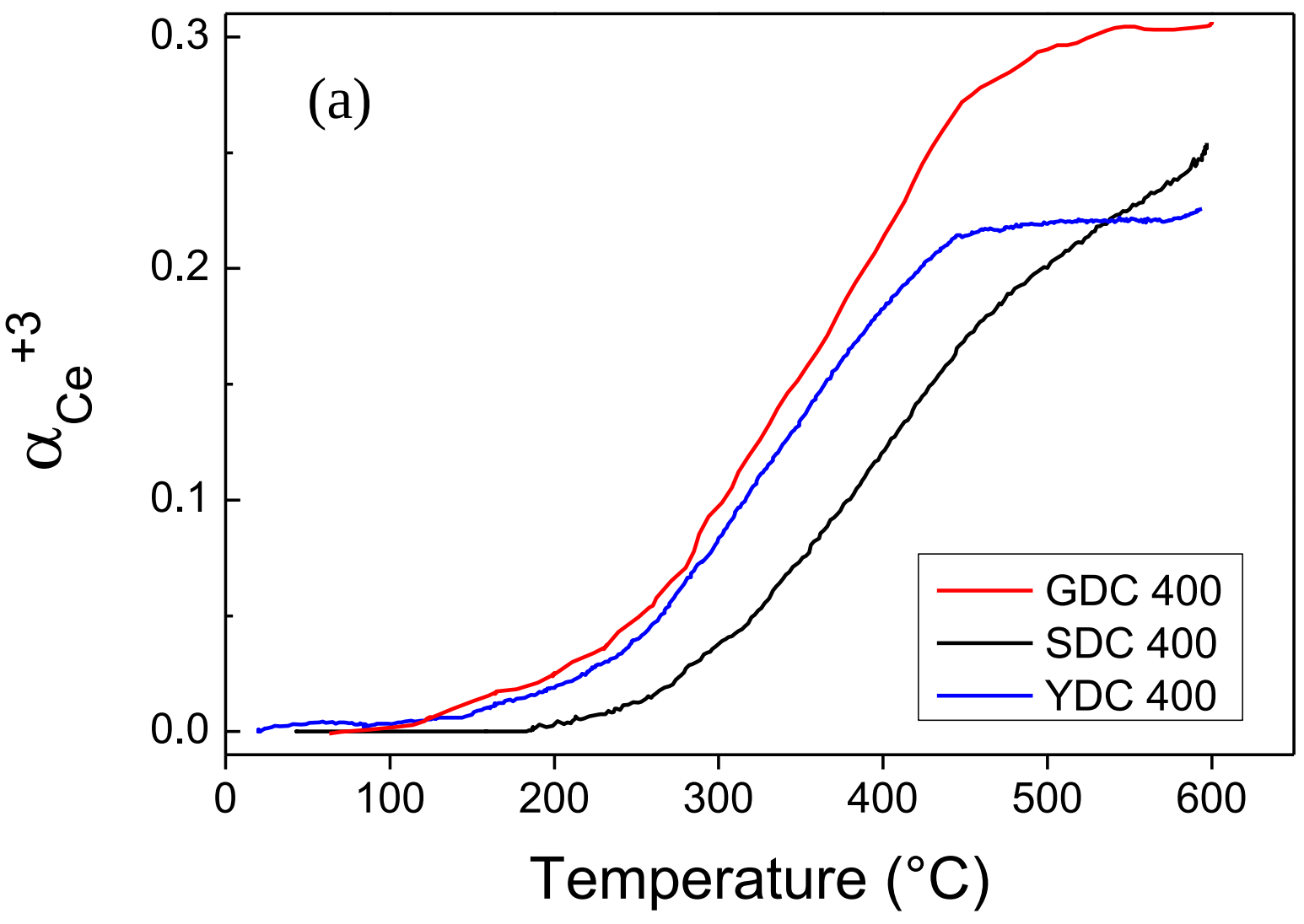

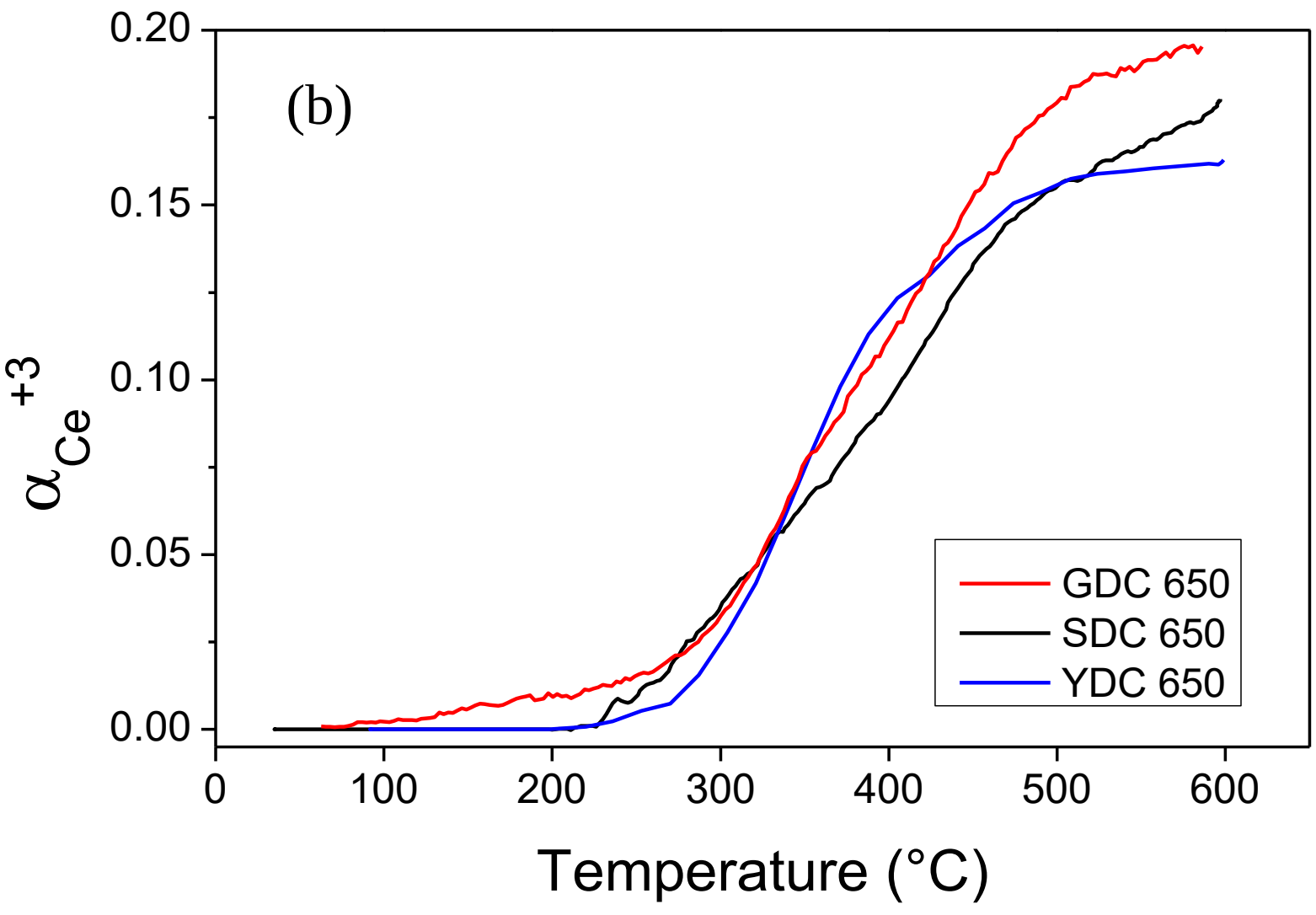

**Figure 9:** $Ce^{3+}$ fraction ($\alpha_{Ce^{3+}}$) for the GDC, SDC and YDC samples, treated at (a) 400ºC and (b) 650ºC, obtained from DXAS analysis.

## Conclusions

In this work we analyzed by the effect of average crystallite size and dopant oxide on the reducibility of $CeO_2$-based nanopowders by temperature-programmed reduction and dispersive X-ray absorption spectroscopy techniques.

We obtained samples formed by single-crystalline particles of controlled grain size, ranging between 4 to around 50 nm of diameter. The addition of $Gd_2O_3$ or $Y_2O_3$ has shown to inhibit grain growth more efficiently than $Sm_2O_3$.

Our results obtained from combined TPR and *in situ* DXAS, evidenced that:

- $Gd_2O_3$-doped ceria display better reducibility than $Y_2O_3$- and $Sm_2O_3$-doped ceria.

- Reducibility is enhanced in the samples formed by small sized crystallites.

Even though all the materials studied in this work exhibited good properties and, therefore, are likely to be used as anodes of intermediate-temperature SOFCs, $Gd_2O_3$-doped $CeO_2$ nanopowders presented the best performance and are more promising for applications.

## Acknowledgements

The present work was partially supported by the Brazilian Synchrotron Light Laboratory (LNLS, Brazil, proposals DXAS-10900, XAFS1-13662 and XAFS1-15360), Agencia Nacional de Promoción Científica y Tecnológica (Argentina, PICT 2015 No. 3411 and PICT 2016 No. 1921) and CAPES-MinCyT bilateral cooperation. The authors thank Prof. Márcia Fantini for her help during X-ray powder diffraction and DXAS measurements and for her valuable comments.

[19]R. Bacani, L. M. Toscani, T. S. Martins, M. C. A. Fantini, D. G. Lamas, S. A. Larrondo. Ceramics International 43, 7851-7860 (2017)
[20]Pei-Lin Chen and i-Wei Chen. J. Am. Ceram. Soc. 79, 1793-1800 (1996).
[21] Hideaki Inaba, Toshifumi Nakajima, Hiroaki Tagawa.Sintering behaviors of ceria and gadolinia-doped ceria. Solid State Ionics106, 263-268 (1998).
[22] W.D. Kingery, H.K. Bowen, D.R. Uhlmann, Introduction to Ceramics, 2nd ed., John Wiley, New York, 1976, p. 456.